\documentclass[aapm,graphicx]{revtex4-1}

\draft 

\usepackage[mathlines]{lineno}
\modulolinenumbers[5]

\usepackage{amssymb}
\usepackage[pdftex]{graphicx}
\usepackage{multirow}
\usepackage{color,soul}

\begin{document}

\title{From conception to clinical trial: IViST - the first multi-sensor-based platform for real-time In Vivo dosimetry and Source Tracking in HDR brachytherapy}

\author{Haydee M. Linares Rosales}
\author{Audrey Cantin}
\author{Sylviane Aubin}
\author{Sam Beddar}
\author{Luc Beaulieu}

\affiliation{$^{1}$D\'epartement de physique, de g\'enie physique et d'optique et Centre de recherche sur le cancer, Universit\'e Laval, Qu\'ebec, Canada. }

\affiliation{$^{2}$D\'epartement de radio-oncologie et Axe Oncologie du CRCHU de Qu\'ebec, CHU de Qu\'ebec - Universit\'e Laval, QC, Canada.}

\affiliation{$^{3}$Department of Radiation Physics, The University of Texas MD Anderson Cancer Center, Houston,TX, United States.}

\affiliation{$^{4}$The University of Texas MD Anderson UTHealth Graduate School of Biomedical Sciences, Houston, TX, United States.}

\email[Corresponding author: Haydee M. Linares Rosales, ]{haydee8906@gmail.com}
\date{}

\begin{abstract}
This study aims to introduce IViST (In Vivo Source Tracking), a novel multi-sensors dosimetry platform for real-time treatment monitoring in HDR brachytherapy. IViST is a platform that comprises 3 parts: 1) an optimized and characterized multi-point plastic scintillator dosimeter (3 points mPSD; using BCF-60, BCF-12, and BCF-10 scintillators), 2) a compact assembly of photomultiplier tubes (PMTs) coupled to dichroic mirrors and filters for high-sensitivity scintillation light collection, and 3) a Python-based graphical user interface used for system management and signal processing. IViST can simultaneously measure dose, triangulate source position, and measure dwell time. By making 100 000 measurements/s, IViST samples enough data to quickly perform key QA/QC tasks such as identifying wrong individual dwell time or interchanged transfer tubes. By using 3 co-linear sensors and planned information for an implant geometry (from DICOM RT), the platform can also triangulate source position in real-time. A clinical trial is presently on-going using the IViST system. 


\end{abstract}


\maketitle 

\section{Introduction}

In vivo dosimetry (IVD) in brachytherapy (BT) aims to quantify in real-time the agreement between the treatment plan and delivered dose. Presently, BT clinics do not verify their treatments in real-time because: (a) commercial real- time systems have small signal-to-noise ratios (SNR), limited time resolution and large measurements uncertainties; (b) laboratory real-time systems can be cumbersome to operate; (c) detector systems do not employ robust error detection algorithms, large false positive rate, poor confidence in error reporting. Several studies \cite{Fonseca-2017, Cartwright-2010, Hardcastle-MOSkin-2010, Kertzscher-2011, Seymour-2011, Kertzscher-2014, Johansen-2018, Sethi-Doppler-US-2018} have focused on developing detectors and methods for real-time source monitoring in BT. The advantages of using plastic scintillator detectors (PSDs) have been outlined in the literature \cite{Beaulieu-Review-2016, Beaulieu-Beddar-2016}]. Nevertheless, PSDs are affected by stem effect and temperature variations \cite{Wootton-Temperature-2013, Beddar-temp-2012}. The proper optimization of the optical chain combined with the implementation of mathematical methods to correct both dependencies on the detector response, the use of multipoint plastic scintillator detector (mPSD) configuration is suitable for real-time source tracking in high dose rate (HDR) brachytherapy  \cite{Archambault-PSD-Select-2005, Archambault-MathForm-2012, Linares-2019}.

The main goal of this paper is to introduce IViST, as a new multi-sensor dosimetry platform for real-time plan monitoring in HDR brachytherapy as well as the first end-to-end application of IViST into the clinical context.

\section{IViST as a platform }

IViST comprises 3 parts: 1) a 3 points mPSD, 2) a light collection system, and 3) a Python-based graphical user interface (GUI) for system management and signal processing. IViST can simultaneously measure dose, triangulate source position, and measure dwell time. 

\subsection{Dosimetry system}

The optimized dosimetry system consisted of a 1.0 mm-diameter core mPSD (using BCF-60, BCF-12, and BCF-10 scintillators) coupled to 15 m-long fiber-optic cable Eska GH-4001 from Mitsubishi Rayon Co., Ltd. (Tokyo, Japan). The mPSD fiber was connected to a data acquisition system consisting of a compact assembly of photomultiplier tubes (PMTs) coupled with dichroic mirrors and filters to achieve a highly sensitive scintillation light collection \cite{Linares-2019}.  The scintillation light generated inside the mPSD is collected in real-time and decoupled using multiple spectral bands \cite{Archambault-MathForm-2012, Linares-2019}.  The PMTs readout is indenpendant from the irradiation unit. Signal acquisition is made at a rate of 100 kHz using a data acquisition board (DAQ) type DAQ NI USB-6216M Series Multifunction from National Instruments (Austin, USA).

\begin{figure}[h]
\centering
\includegraphics[scale=0.28]{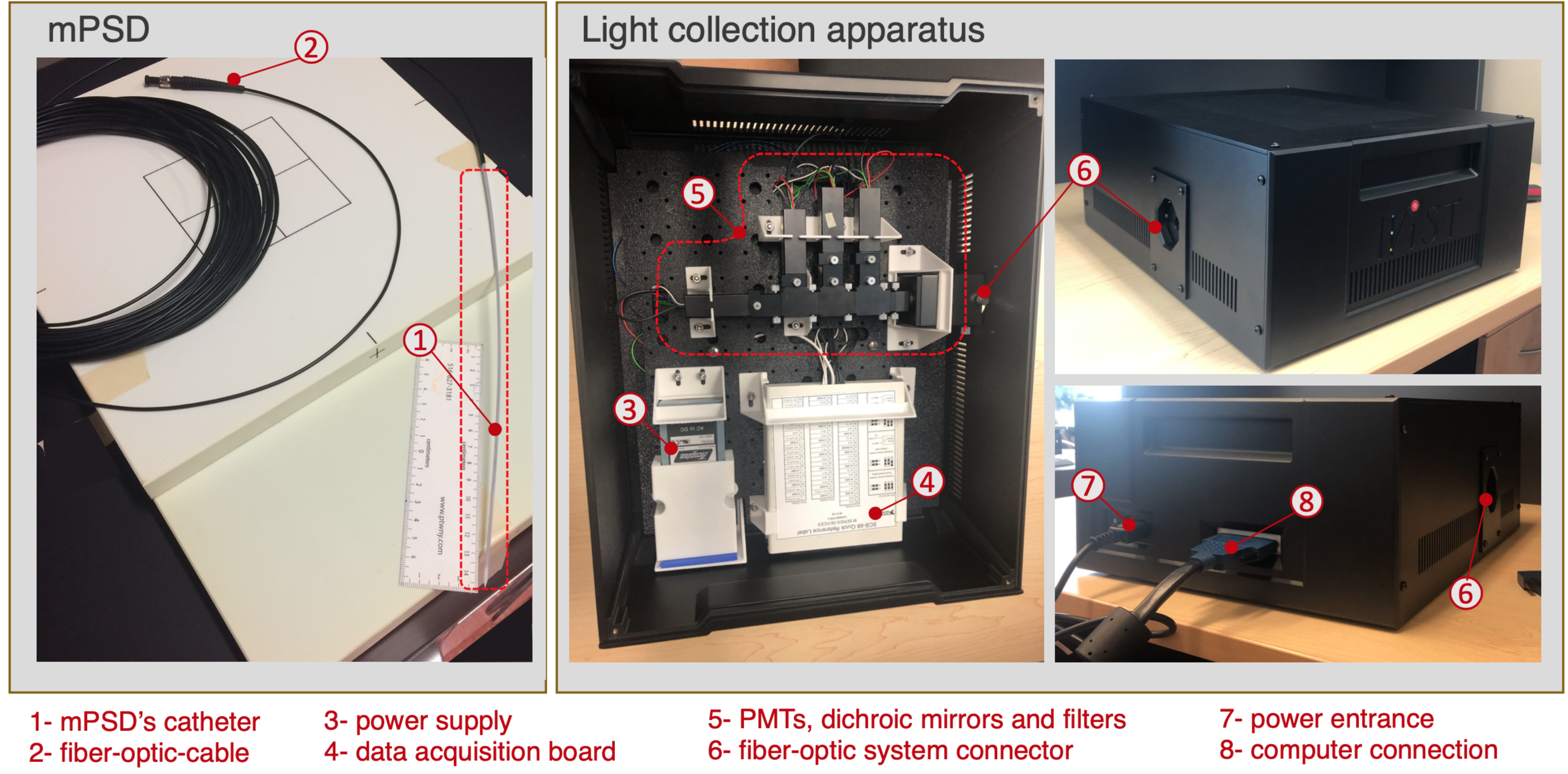}
\caption{\label{IVisT_dosim_system} IViST's dosimetry system. A detailed list of sub-components can be found in Linares Rosales et \textit{al.} \cite{Linares-2019-dosimetry-mPSD} }
\end{figure}

The design of the in vivo dosimetry system is shown in figure \ref{IVisT_dosim_system}. The detector was made light-tight to avoid environmental light. The mPSD can be inserted into a 24-cm plastic catheter (Best Medical International, Springfield, VA, USA). In vivo measurements can be made using a supplemental catheter during HDR brachytherapy. The light collection apparatus was enclosed into a custom-made box containing the light acquisition system as well as the DAQ. It is a passive light collection system that does not represent any electrical risk. The system's dosimetric performance have been evaluated in HDR brachytherapy, covering a range of 10 cm of source movement around the sensors \cite{Linares-2019, Linares-2019-dosimetry-mPSD}. Within a range of distances of 6 cm from the source, IViST can track the $^{192}$Ir  source with at least a 1 mm positional accuracy, and dose agreement with TG43-U1 \cite{TG-43-Update} expected dose within 5 \% \cite{Linares-2019-dosimetry-mPSD}. The detector further exhibited no angular dependence and the system high collection rate allows for dwell time assessment with a maximum average weighted deviation of 0.33 $\pm$ 0.37 s \cite{Linares-2019-dosimetry-mPSD}.

\subsection{Graphical user interface for dosimetry system control}
\label{GUI_section}

\begin{figure}
\centering
\includegraphics[scale=0.40]{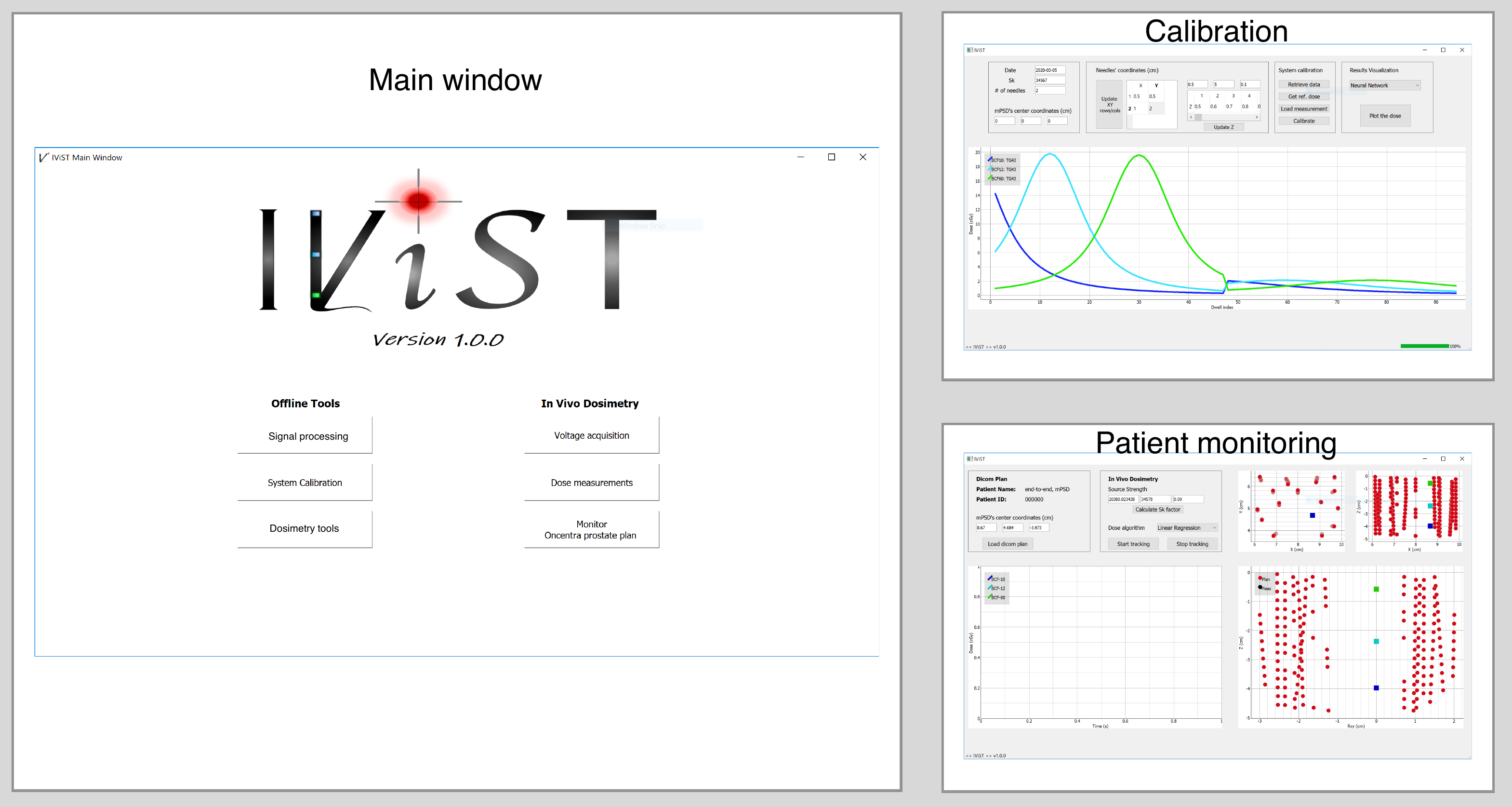}
\caption{\label{IViST_win} IViST’s graphical user interface. Illustration of some of the available functionalities.}
\end{figure}

\begin{figure}
\centering
\includegraphics[scale=0.38]{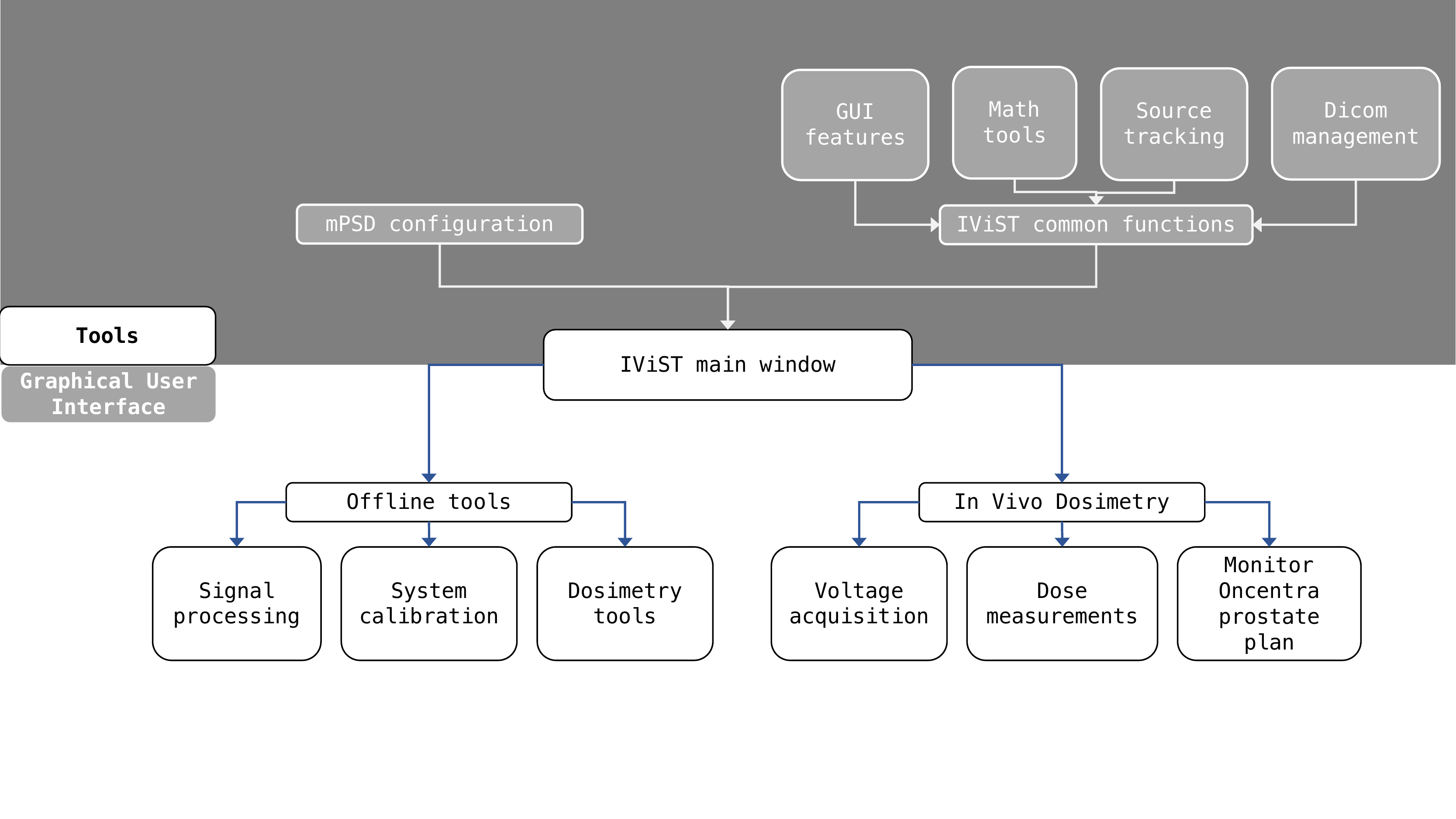}
\caption{\label{IViST_mod} IViST's code's diagrams.}
\end{figure}

Figure \ref{IViST_win} illustrates the graphical user interface (GUI) of the key software tools available to control the dosimetry system. The GUI was designed for simple operation by any users. PyQt5 was used to create the GUI. Figure \ref{IViST_mod} summarizes the main functionalities of the GUI and the relationship among them. The software’s main windows give the user access to 2 main modes: Offline and In vivo.

Within \textbf{Offline tools}:
\begin{itemize}
    \item \textbf{Signal processing}: A measurement file is loaded in this module by the user. Then, the module allows for easy extraction and manipulation of the raw data previously collected. It can also be used to extract the statistics associated with each dwell position (e.g. measured dwell time, mean raw signal, mean dose per dwell).
    
    \item \textbf{System calibration}: The module receives a measurement data set, and based on TG43’s expected dose, generates the calibration models for raw signal translation into a dose value. Among the main features of this module, it allows for dose prediction with the hyper-spectral approach proposed in \cite{Archambault-MathForm-2012} as well as a Random Forest algorithm and Neural Network.
    
    \item \textbf{Dosimetry tools}: Performs the dosimetric analysis of the collected data. The user could, for instance, verify the agreement of the measured dose with that expected from TG43 and calculate the source position for the measured plan. The module also includes a catheter shift correction feature.

\end{itemize}

The In Vivo Dosimetry module contains the tools used for real-time data acquisition:
\begin{itemize}
    \item \textbf{Voltage acquisition}: Collects the raw measurement data from each PMT in volts. 
    \item \textbf{Dose measurement} is a similar to the offline module, excepting that a calibration data file has to be available in advance.
    \item \textbf{Monitor Oncentra prostate plan}: Performs real-time tracking of a patient plan created in Oncentra Prostate (Nucletron, Elekta, Sweden). The module reads the plan data contained in the \textit{*.cha} files exported from the TPS, sets the treatment data as reference and monitors the patient plan in real-time. The system should be calibrated in advance. The GUI allows for real-time visualization of the measured dose at the mPSD location as well as the source location.
\end{itemize}

\section{In vivo measurements for a prostate treatment: First Patient Measurements}

In vivo dose monitoring was carried out for the first patient enrolled in an IRB approved protocol. Seventeen treatment catheters were positioned and reconstructed to deliver 15 Gy in a single fraction. An additional catheter devoted to measurements with the mPSD was introduced in the prostate volume. The mPSD’s catheter was also reconstructed, but it was not used during the dose optimization process in the treatment planning system.

\begin{figure}
\centering
\includegraphics[scale=0.30]{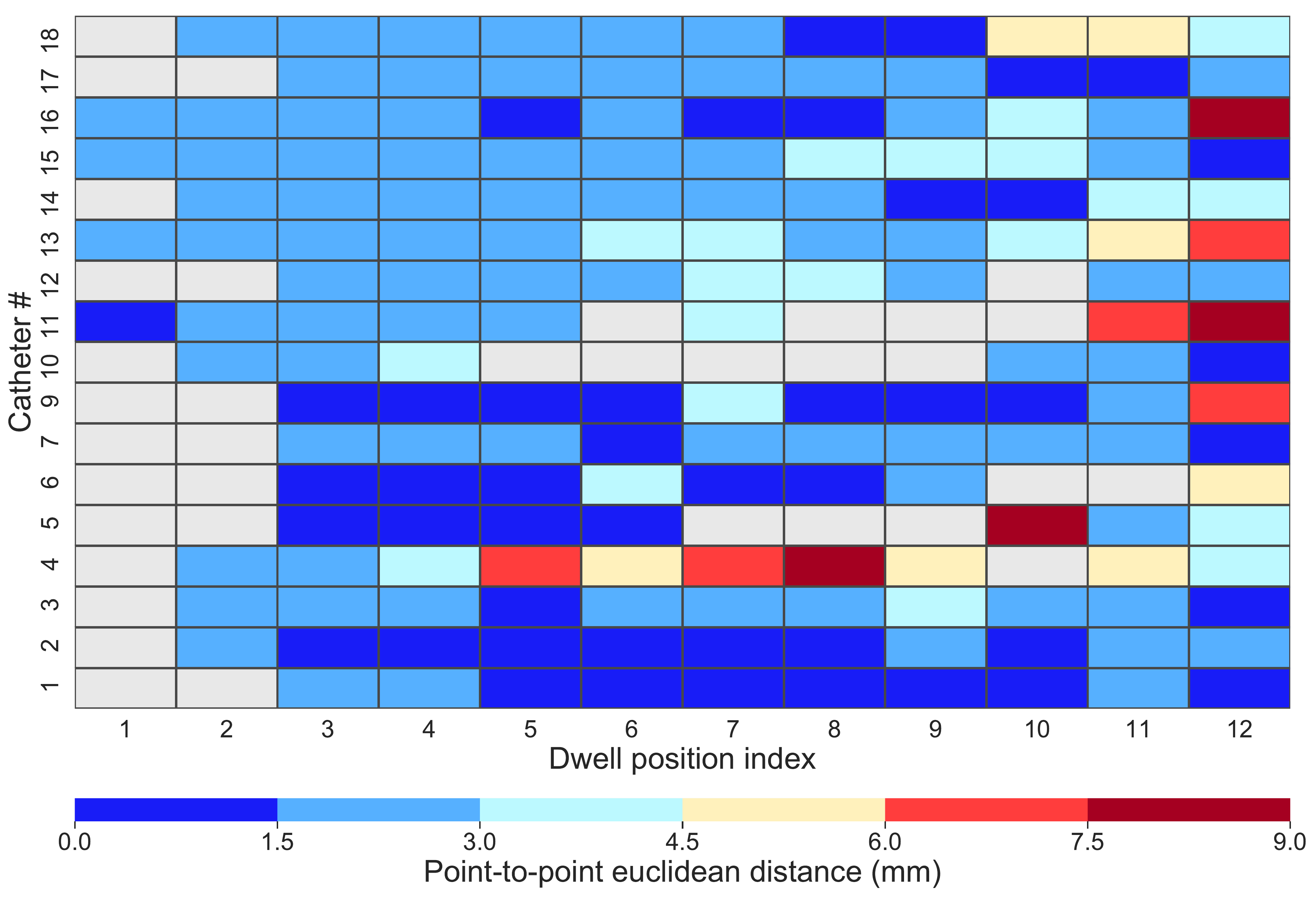}
\caption{\label{Prost_1_heatmap} Source position triangulation results. Heat map of deviations between the planned position and the triangulated one in a 3D space. The cells in gray correspond to non-programmed dwell positions.}
\end{figure}

\begin{figure}
\centering
\includegraphics[scale=0.34]{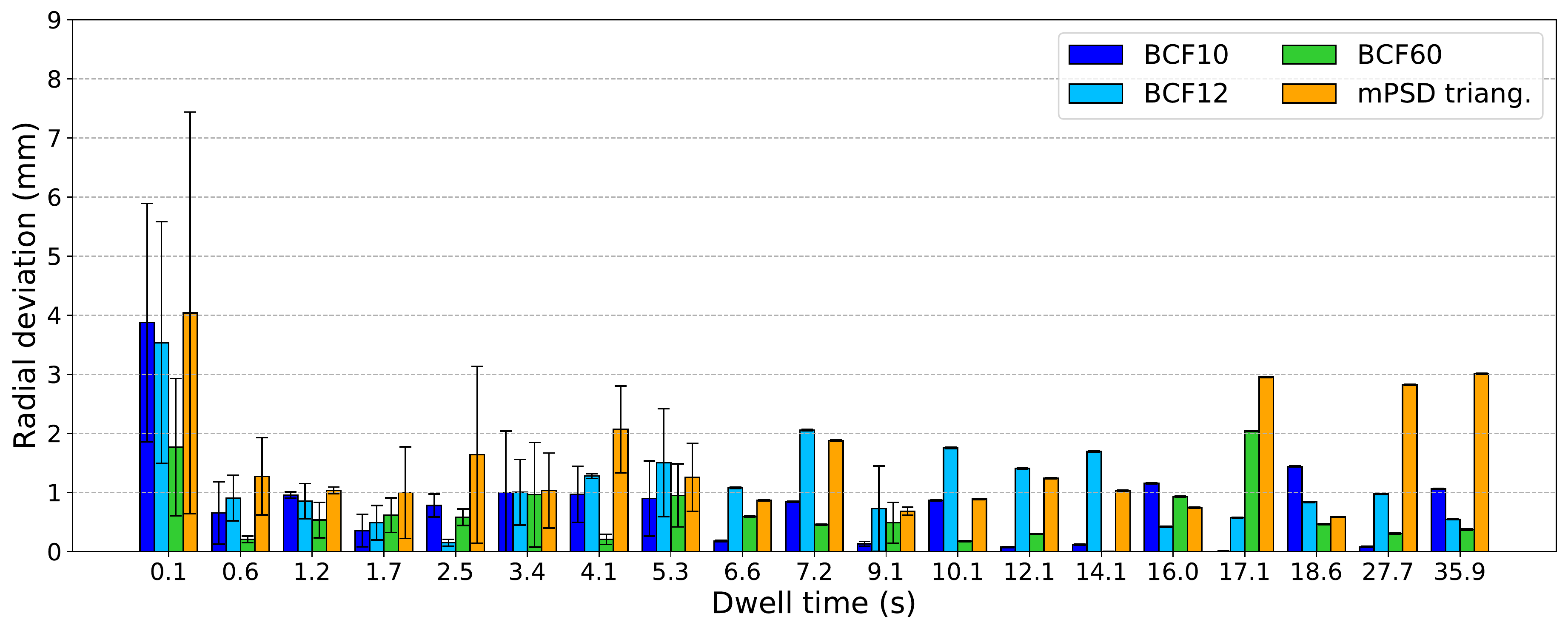}
\caption{\label{Prost_1_R_dwellt} Scintillator deviation of the expected source-to-dosimeter radial deviations as a function of the source planned dwell time.}
\end{figure}

Figure \ref{Prost_1_heatmap} shows the heat map of deviations obtained between the planned source dwell positions and the results from the triangulation with the mPSD. The x-axis corresponds to the dwell position index within each catheter, while the y-axis corresponds to the treatment catheter number. Catheter 8 is not plotted because it corresponds to the catheter dedicated to measurements with the mPSD. In total, 168 source dwell positions were tracked. As could be observed in figure \ref{Prost_1_heatmap}, in most of the cases, the radial deviations remain below 3 mm. However, there are some dwell positions with deviations reaching 9 mm. This effect could be explained by the small source dwell times, 0.1 s (resultant from the dose optimization process), programmed at dwell-positions. Figure \ref{Prost_1_R_dwellt} presents the observed radial deviations as a function of the delivered dwell times for BCF10, BCF12, BCF60 sensors as well as for the mPSD's triangulation results. As could be observed in figure \ref{Prost_1_R_dwellt}, for dwell times of 0.1 s, radial deviations in the order of 6 mm should be expected. IViST’s data is collected at a rate of 100 kHz but averaged each 0.1 s. Thus a dwell time of 0.1 s makes almost impossible to distinguish one single dwell position from the previous/subsequent one, leading to a failed triangulation for those locations. However, for any dwell-time of 0.5 s or more, the weighted average is  2.17 $\pm$ 1.15 mm. 

\begin{figure}
\centering
\includegraphics[scale=0.33]{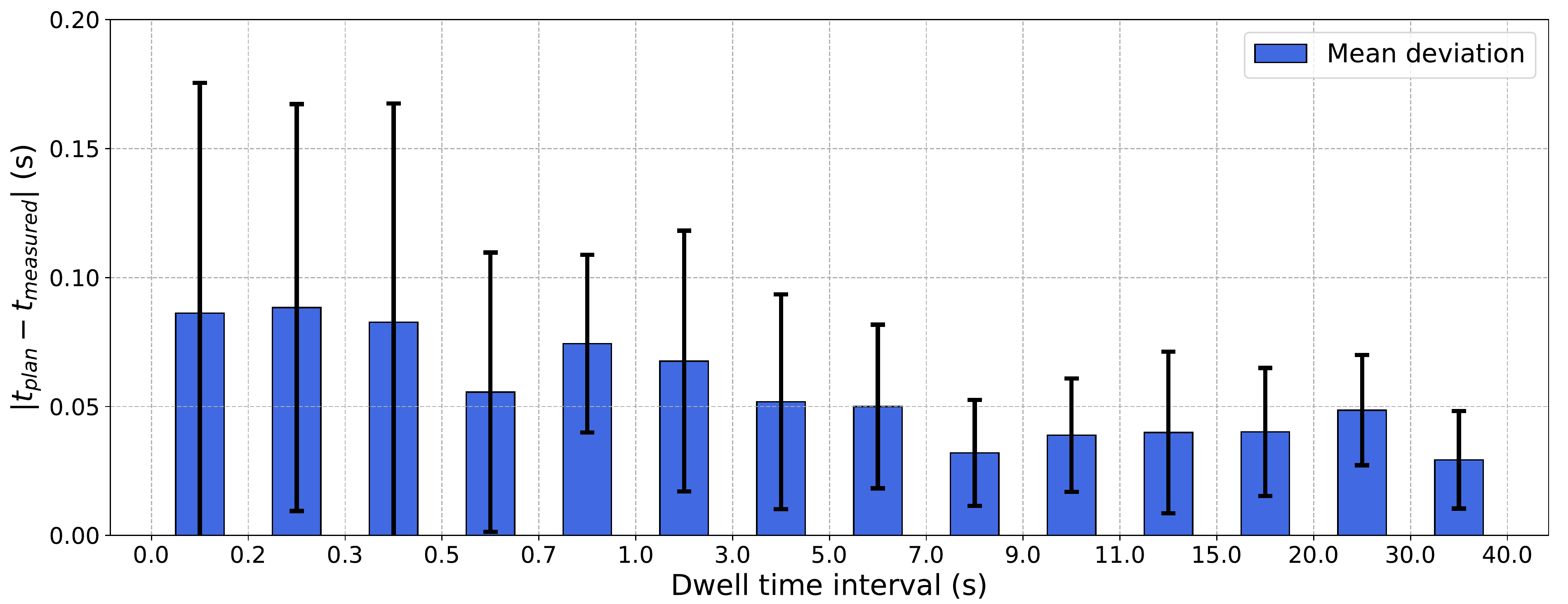}
\caption{\label{Prost_1_R_dwellt_dev} Deviation of mPSD measured dwell times from planned dwell times as function of planned dwell-time intervals.}
\end{figure}

Figure \ref{Prost_1_R_dwellt_dev} shows the mean deviations between the measured dwell times for our mPSD system and the planned ones. The measured dwell times at all the distances to the source were agreement with the planned dwell time to better than 0.17 s. The overall average difference obtained was 0.06$\pm$0.04 s. Larger deviations were observed for dwell positions with dwell times inferior to 0.5 s, especially for those positions at long distances from the scintillators' effective volumes. Even though the longest mPSD-to-source distance measured for this clinical case was 55 mm, the high gradient field characteristic of the $^{192}$Ir source allowed to us obtain a sharp pulse of signal and, as a consequence, proper differentiation of the signal from one position to the subsequent one. A similar dwell time analysis was performed by Linares Rosales \cite{Linares-2019-dosimetry-mPSD} et \textit{al.} through in-water measurements during HDR brachytherapy. They reported a maximum deviation of 0.33 $\pm$ 0.37 s for a range of mPSD-to-source distance up to 10 cm. However, for the same range of distance to the explored in this paper, the results are equivalents \cite{Linares-2019-dosimetry-mPSD}. 

\section{Conclusions}
A novel, multi-sensor dosimetric system was optimized for HDR brachytherapy in vivo dosimetry and enabled measurements over a wide range of clinically relevant dose rate. The system presented allows real-time dose, source position and dwell-time measurements. It has numerous potential in vivo QA/QC applications beyond the currently available commercial dosimeters and build on its widely-known intrinsic energy independence and water equivalence properties. The first clinical in vivo case measured with iViST has been presented here and the clinical trial is on-going.

\section*{Acknowledgements}

This work was supported by the National Sciences and Engineering Research Council of Canada (NSERC) via the NSERC-Elekta Industrial Research Chair grant \# 484144-15 and RGPIN-2019-05038. Haydee Maria Linares Rosales also acknowledges support from Fonds de Recherche du Quebec - Nature et Technologies (FRQ-NT).

\end{document}